\documentclass[letterpaper,preprint]{revtex4}
\usepackage{graphicx}
\usepackage{subfigure}
\begin{document}
\title{Population inversion in hyperfine states of Rb with a single nanosecond chirped pulse in the framework of a 4-level system}

\author{G. Liu, V. Zakharov, T. Collins, P. Gould$^\dag$ and S.A. Malinovskaya\\ Department of Physics and Engineering Physics, Stevens Institute of Technology, Hoboken, New Jersey 07030, USA\\
$^\dag$ Department of Physics, University of Connecticut, Storrs, Connecticut 06269, USA}

\begin{abstract}

We implement a 4-level semiclassical model of a single pulse interacting with the hyperfine structure in ultracold rubidium aimed at control of population dynamics and quantum state preparation. We discuss a method based on pulse chirping to achieve population inversion between hyperfine states of the 5S shell. The results may prove useful for quantum operations with ultracold atoms.
\end{abstract}
\maketitle

Ultracold atoms, with their minimal spectral broadening and long interaction times, are ideal systems to explore and implement quantum state preparation and manipulation [1]. In particular, Raman processes in ultracold systems have proven to be a very useful tool in areas ranging from laser cooling and atom optics [2] to quantum information processing [3] and ultracold molecules [4]. The standard method of driving a stimulated Raman transition between two levels is to apply two beams of light, separated in frequency by the two level splitting [5]. Here we propose to use instead a single pulse which is frequency chirped trough the Raman resonance. This has the advantage of simplicity, high speed, and potentially efficient transfer. Specifically, we examine Raman transitions between the ground-state hyperfine levels of Rb upon interaction with a single linearly polarized and linearly chirped laser pulse. With chirped light, the excited-state hyperfine structure does not generally allow the realization of the ideal three-level system often assumed in Raman transitions. We deal with this complication by treating the atom as a 4-level system.

The linearly chirped pulse is described analytically as
\begin{equation}
E(t)=E_{0}(t)*\cos[2 \pi \omega_{0}(t-T)+\alpha(t-T)^{2}/2]\label{field}
\end{equation}
Here $E_0(t)$ is the Gaussian envelope of the form $E_0 \exp(-(t-T)^2/\tau_0^2)$, T is the time of the pulse peak intensity, $\tau_0$ is the pulse duration related to the full width at the half maximum (FWHM) of the field E(t) as FWHM=$\tau_0 2\sqrt{ \ln2}$, $\omega_0$ is the carrier frequency of the pulse at the peak value of the field, and $\alpha/2\pi$ is the linear chirp (GHz/ns).
The atoms are considered to be in an ultracold dilute gas, thus
eliminating Doppler shifting of the laser field in the atomic
frame. Such ultracold gasses of atoms may be readily produced and
contained in a laboratory setting using a magneto-optical trap (MOT) \cite{Ra87}. MOTs have been used to create ultracold gasses and perform, e.g., spectroscopy \cite{Ma07,Ye08} and quantum
state manipulation \cite{Dj06}.

A semiclassical model is used to describe the interaction of a linearly chirped pulse
with the hyperfine (HpF) structure of the ultracold Rb.
 The two-photon Raman transitions are investigated
aiming at achieving full population transfer from the lower to the upper HpF state
of the $5S$ orbital with the aid of the energetically higher $5P_{1/2}$ or $5P_{3/2}$
states. As an initial condition, the electron resides in the ground HpF
state of the $5S$ orbital. Our goal is to determine the proper field
parameters that may accomplish population inversion within
the $5S$ orbital by taking advantage of the four optically allowed transitions
that exist within both the D1 and D2 lines of $^{85}$Rb and $^{87}$Rb.

\begin{figure}
\centerline{\includegraphics[width=5in]{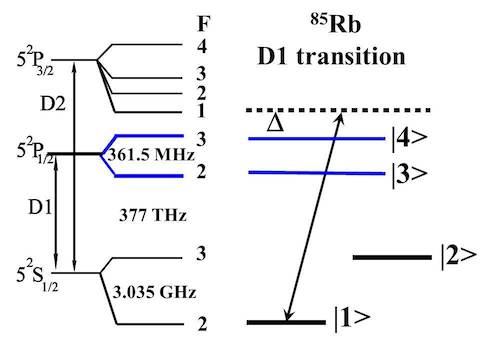}}
\caption{ Four optically attainable hyperfine states of $5S$ and $5P$ shells with the energy differences that correspond to the D1 line, modeled by a 4-level system. Initially, the population is in the ground state $\left|1\right\rangle $. Note that the hyperfine splitting of the $5S_{1/2}$ orbital is approximately an order of magnitude greater than the splitting of $5P_{1/2}$ orbital. }\label{4-level system}
\end{figure}

Such an analysis has been done before using a model of three-level $\Lambda$ system which included
two hyperfine levels of the $5S$ shell and the lowest hyperfine state of the $P$ shell, \cite{Co-1-12}. It revealed that
the population inversion within the $5S$ orbital was attainable using chirped, nanosecond laser
pulses of beam intensities on the order of $1\frac{kW}{cm^{2}}$ and one-photon detuning on the order of the $5S$ hyperfine splitting. In this paper, a model of the 4-level system is used to address all optically allowed transitions between hyperfine states belonging to $5^{2}S_{1/2}$ and $5^{2}P_{1/2}$ or $5^{2}P_{3/2}$ states; the transitions are optically allowed between those HpF states whose total angular
momentum quantum number $F$ changes as $\Delta F=\{1,0,-1\}$. For, e.g., $^{85}$Rb, these are two hyperfine states having F=2,3, separated by $\omega_{21}$. Note, that the splitting $\omega_{43}$ between F=2 and F=3 states in the 5P shell is an order of magnitude less than that between the states with the same quantum numbers in the 5S shell, see Fig. (\ref{4-level system}) and \cite{key-1}. Because of that, as our calculations show, a pulse having a duration on the order of $\omega_{21}^{-1}$ and a chirp rate satisfying the adiabaticity condition, populates the upper states $|3>$ and $|4>$ almost identically, demonstrating that the 4-level system behaves as an effective three-level $\Lambda$ system. However, for longer pulses the resulting difference in the population of the upper hyperfine states may be significant. Thus, the model of the  4-level system is more general and gives a more complete picture of population distribution for the field parameters associated with higher spectral resolution. One of the main points of the present work is to analyze how including the second excited state affects the Raman transfer. This is important for experimental implementations since this fourth level cannot in general be eliminated.

A complete treatment would require inclusion of all the magnetic sublevels of each hyperfine level, but in an experiment, the system could be restricted to four states by preparing a single state via optical pumping and using selection rules based on polarization of the chirped light to restrict the number of states involved.

The field-interaction Hamiltonian used in the time-dependent Schr$\ddot{o}$dinger equation reads
\begin{equation}
\hat{H}=h \left[\begin{array}{cccc}
\Delta+\omega_{43}+\alpha(t-T) & 0 & \Omega_{R}(t)/2 & \Omega_{R}(t)/2\\
0 & \Delta+\omega_{43}+\omega_{21}+\alpha(t-T) & \Omega_{R}(t)/2 & \Omega_{R}(t)/2\\
\Omega_{R}(t)/2 & \Omega_{R}(t)/2 & 0 & 0\\
\Omega_{R}(t)/2 & \Omega_{R}(t)/2 & 0 & \omega_{43}\end{array}\right]\end{equation}
Here $\Omega_{R}(t)\equiv - \mu E_{0}(t)/h$ is the Rabi frequency with the peak value $\Omega_R$. If the one-photon detuning $\Delta$ is equal to zero, the laser carrier frequency becomes equal to the $\omega_{41}$ transition frequency at the peak value of the field amplitude.  The value of the transition dipole moment $\mu$ for $^{87}Rb$ and $^{85}Rb$ is $3.58 \cdot10^{-29}$ Cm for the D2 line, and $2.54 \cdot 10^{-29}$ Cm  for the D1 line.  In our calculations, transition probabilities between hyperfine levels within these lines are considered to be equal. Besides, we ignore spontaneous emission because of our fast time scales.

We investigated numerically the population transfer within the 4-level system as a function of the pulse duration $\tau_0$,  the linear chirp rate $\alpha/2\pi$, the one-photon detuning $\Delta$, (where $\Delta=\omega_0 - \omega_{41}$), and the peak Rabi frequency $\Omega_R$. Intuitively, we expect that a negative chirp should be implemented in order to perform the two-photon population transfer with a narrow-band pulse by sweeping through the one-photon resonances beginning from the high-frequency resonance with the $|1>-|4>$ transition. The peak Rabi frequency was chosen to be 30.35 GHz (10$\omega_{21}$) and 3.035 GHz ($\omega_{21}$) corresponding to the peak value of the field intensity about $83$ kW/cm$^2$ and $833.8$ W/cm$^2$, respectively.

\begin{figure}
\vspace{20pt} \centerline{
\includegraphics[width=13.cm]{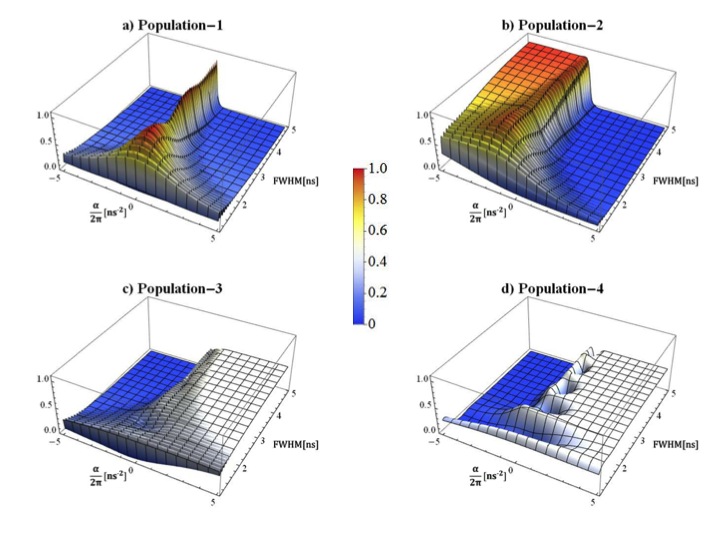}}
\caption{Final population distribution in the 4-level system as a function of $\alpha/2\pi$ and FWHM for $\Omega_R$=30.35 GHz, and $\Delta = 0$.
 The values of the system parameters
are  $\omega_{21}$ = 3.035 GHz, $\omega_{43}$ = 0.362 GHz, characteristic for $^{85}$Rb \cite{key-1}, the peak Rabi frequency is . }\label{fourpopullinearch}
\end{figure}
The results of calculations for $\Omega_R$ = 30.35 GHz (10$\omega_{21}$) are shown in Fig.(\ref{fourpopullinearch}). An informative region is presented by the FWHM from 1.0 ns to 5.0 ns and the $\alpha/2\pi$ from zero to $\pm 5$ GHz/ns. These ranges of parameters may be reached using a setup similar to that  described in \cite{Go07}.
 The figure shows that for pulses having $\tau_0 > \omega_{21}^{-1}$ (FWHM = 0.56 ns) and a chirp rate such that $ |\alpha/(2\pi) | \tau_0 \simeq \omega_{21}$, the solution is a  Rabi type oscillation as a function of $\tau_0$  and $\alpha/(2\pi)$   resulting in a periodic population transfer to the upper hyperfine state $|2>$ of $5S$ shell. For $ |\alpha/(2\pi) | \tau_0 > \omega_{21}$, the dynamics follows the adiabatic regime.

 For small values of FWHM, such that the pulse spectral width is $1/\tau_0 > \omega_{21}$, the population dynamics follows the pulse area solution. For such broad pulses, chirping is of a secondary importance since population inversion may be achieved following the pulse area solution.

\begin{figure}
\vspace{20pt} \centerline{
\includegraphics[width=13.cm]{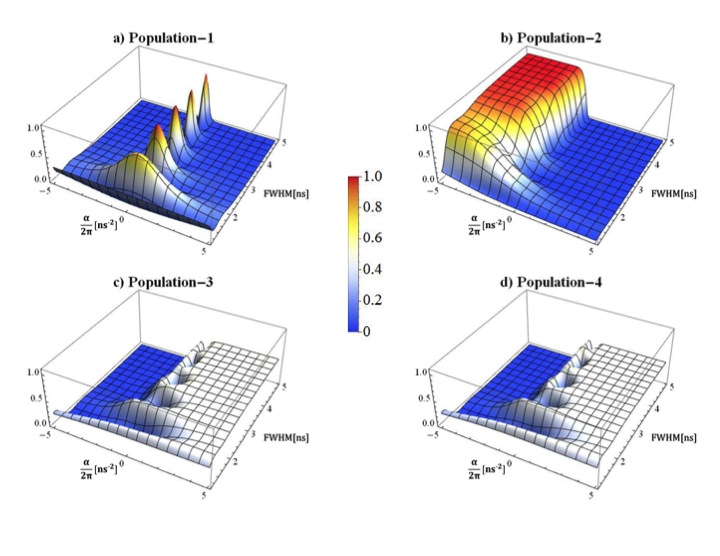}}
\caption{ Final population distribution in the 4-level system  as a function of $\alpha/2\pi$ and FWHM for $\Omega_R$=3.035 GHz, and $\Delta = 0$. The values of the system parameters
are  $\omega_{21}$ = 3.035 GHz, $\omega_{43}$ = 0.362 GHz, characteristic for $^{85}$Rb \cite{key-1}. }\label{4lvl-Rabi1}
\end{figure}

\begin{figure}
\centerline{
\includegraphics[width=3.7in]{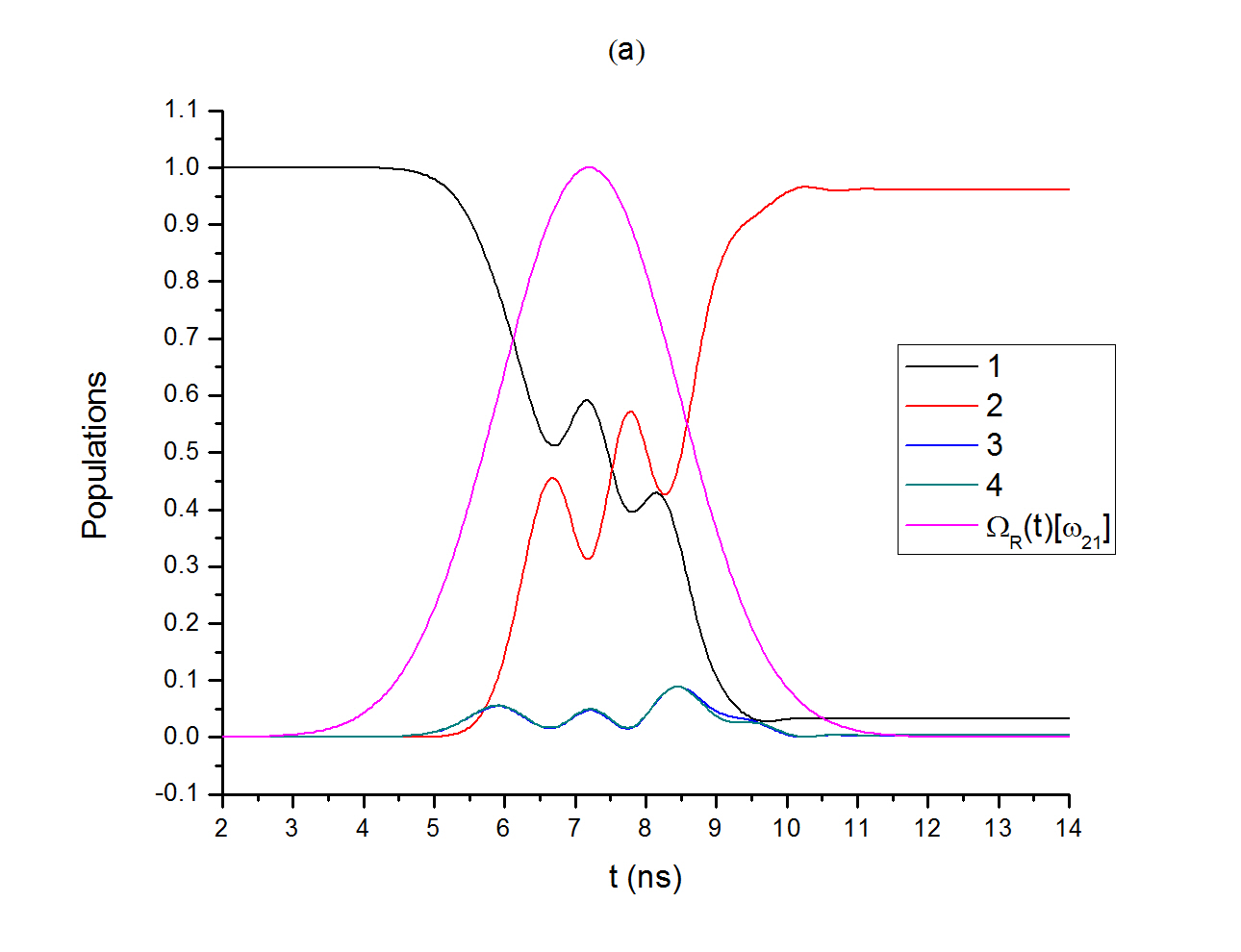}\hspace{0.5cm}\includegraphics[width=3.7in]{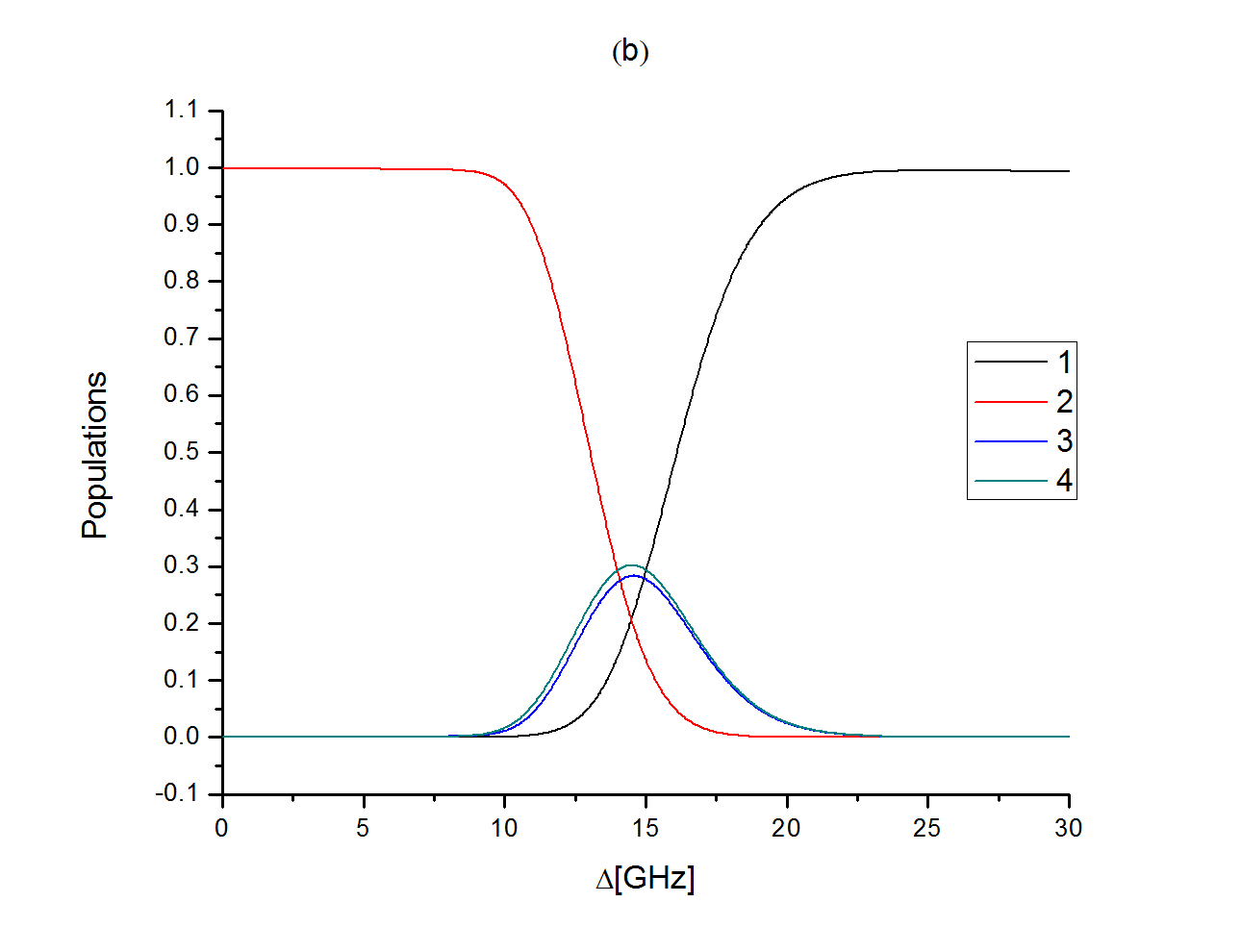}}
\caption{ (a) Time-dependent picture of the population dynamics of four hyperfine states leading to population transfer to the final state $|2>$. The field parameters are $\Omega_R$=3.035 GHz, FWHM=2.99353 ns and $\alpha/(2\pi)$=-2.94752 GHz/ns. (b)  The end-of-pulse population of four hyperfine states as a function of the one-photon detuning $\Delta$. The field parameters are $\Omega_R$=3.035 GHz, $\alpha/2\pi$=-1 GHz/ns and FWHM=17 ns. }    \label{adiab}
\end{figure}

When the condition $ |\alpha/(2\pi) | \tau_0 > \omega_{21}$ is satisfied,
a smooth, adiabatic regime of population transfer to state $|2>$ may be achieved by implementing a lower intensity field satisfying the Landau-Zener adiabaticity condition $|\alpha/(2\pi)|<< \Omega_R^2$. If we consider the peak intensity  value  of 833.8 W/cm$^2$, which gives $\Omega_R=\omega_{21}=3.035$ GHz,  a significant  population transfer to the final state $|2>$ is manifested starting from FWHM$\approx$2.5 ns ($\tau_0$=1.5 ns) and $\alpha/(2\pi)$ = -2 GHz/ns, giving $\alpha/(2\pi) \tau_0 \approx$3 GHz; and the adiabatic regime is reached for FWHM$\approx$3 ns and $\alpha/(2\pi)\approx$-3 GHz/ns, see Fig.(\ref{4lvl-Rabi1}).
In the time-dependent picture, Fig(\ref{adiab},a), as the pulse instantaneous carrier frequency gradually decreases toward the first one-photon resonance with $\omega_{41}$, the population adiabatically moves from the initial state $|1>$ to the final state $|2>$ through the excited state manifold. In the vicinity of the peak value of the field intensity and $\omega_0 \approx \omega_{41}$, almost half of the population has been transferred from the initial to the final state and about 1$\%$ resides in states $|3>$ and $|4>$. As the instantaneous frequency further decreases  toward the second one-photon resonance with the transition frequency $\omega_{42}$, more population is accumulated in the final state $|2>$, resulting in almost complete population transfer to this state by the end of the pulse. The mechanism relies on two-photon adiabatic passage: As the frequency moves toward the first resonance, the excited states work as mediators in the off-resonant two-photon transition to the final state $|2>$. When a gradual decrease in the instantaneous frequency leads to the second one-photon resonance with $\omega_{42}$, it accomplishes the two-photon resonance condition and effectively moves population from the excited states to the final state. The effective detuning switches sign at this point. The rest of the pulse completes population transfer to the final state off-resonantly. A one-photon detuning on the order of the peak Rabi frequency leads to similar population dynamics with adiabatic passage to the final state,  Fig(\ref{adiab},b). Further  increase of the one-photon detuning diminishes the contribution of the excited state manifold and prevents the adiabatic population flow to the final state for $\Delta >> \Omega_R$.

\begin{figure}
\vspace{20pt} \centerline{
\includegraphics[width=3.7in]{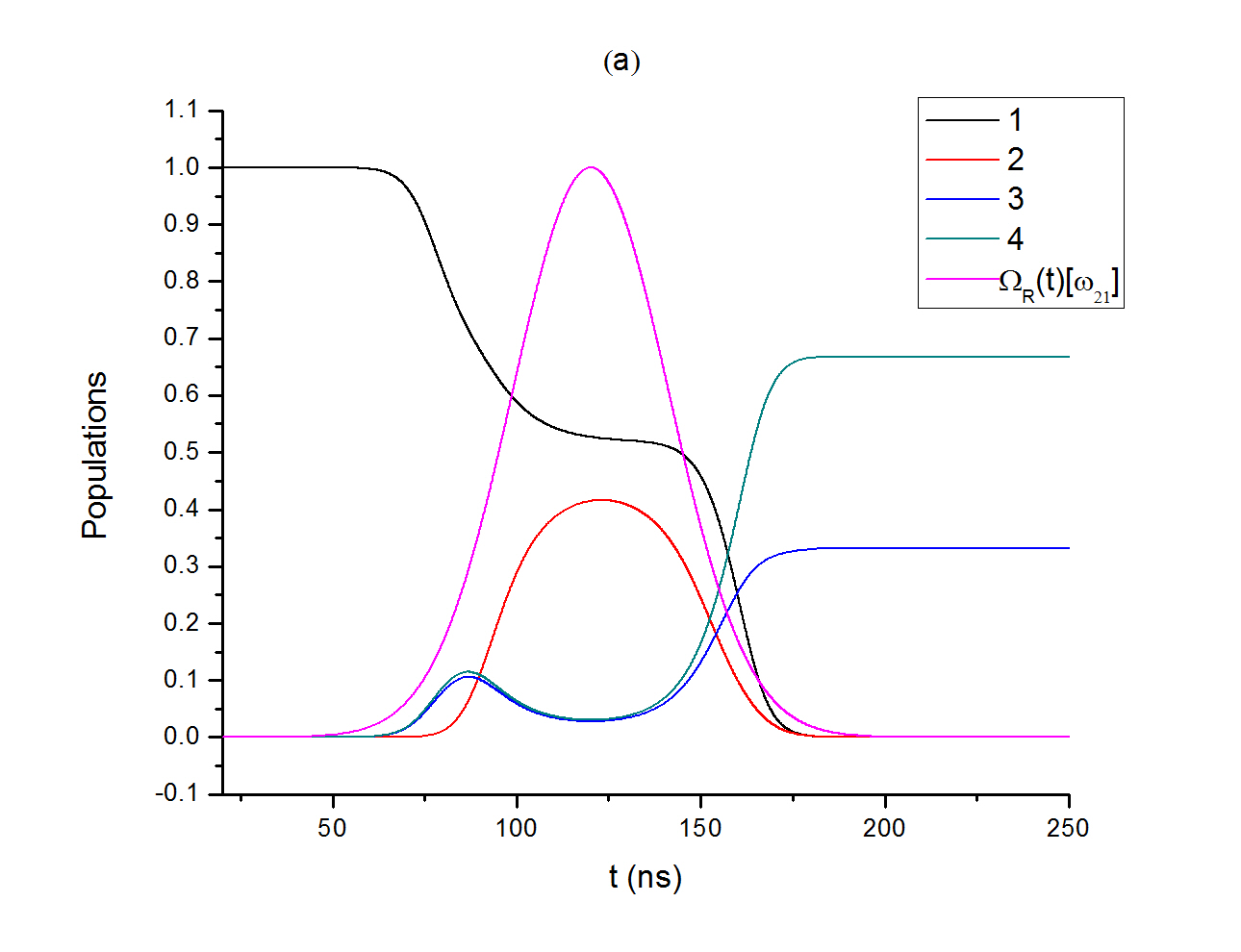}\hspace{0.5cm}\includegraphics[width=3.7in]{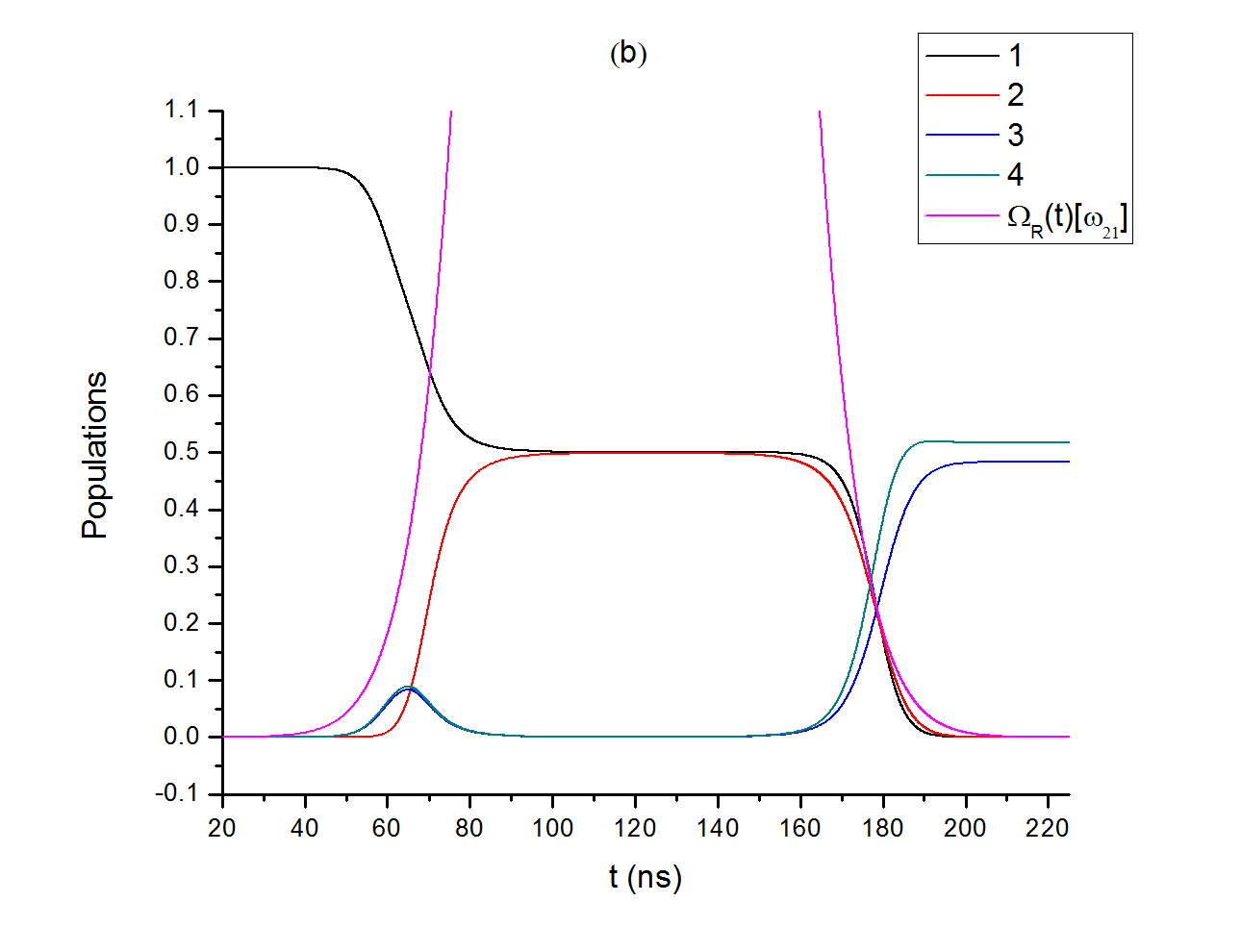}}
\caption{(a,b) Time-dependent picture of adiabatic passage to states $|3>$ and $|4>$ achieved by a pulse having FWHM=50 ns, $\alpha/2\pi$ = -0.026 GHz/ns and (a) $\Omega_R$=3.035 GHz, (b) $\Omega_R$=30.35 GHz.}\label{small_ch_FWHM_depnd}
\end{figure}

When a pulse of much longer duration and small negative chirp rate is applied, such that $ |\alpha | \tau_0 << \omega_{21}$, no transfer to state $|2>$ is observed  regardless of the field intensity; see Fig.(\ref{small_ch_FWHM_depnd}a,b) where the dependence of the state population is shown as a function of time for a small negative chirp rate $\alpha/2\pi$=-0.026 GHz/ns, FWHM=50 ns and  $\Omega_R$=3.035 GHz, (a), and $\Omega_R$=30.35 GHz, (b). Here adiabatic passage takes place to the excited state manifold leading to differently populated states $|3>$ and $|4>$.

For positive values of the chirp rate and $\tau_0 > \omega_{21}^{-1}$, the excited state manifold is populated adiabatically in Figs.(\ref{fourpopullinearch},\ref{4lvl-Rabi1}) owing to the resonance condition being satisfied with these closely spaced states at the peak field intensity and subsequent moving away from all resonances in the 4-level system.

In summary, we have demonstrated that a single chirped nanosecond pulse of intensity on the order of 800 W/cm$^2$
may be successfully implemented to manipulate valence electron dynamics in alkali atoms at ultracold temperatures aiming at the population transfer within the hyperfine structure. The results are based on the  developed model of a classical pulse interaction with the 4-level system representing optically accessible hyperfine states in ultracold $^{85}$Rb. Numerical analysis is performed revealing the dependence of population dynamics and the quantum yield on the key field parameters: the pulse duration, the linear chirp rate, the peak amplitude and the one-photon detuning. We analyzed a possibility to perform the population inversion within the hyperfine states of the $5S$ shell. The adiabatic region of light-matter interaction leading to population inversion was found for parameters that satisfy the adiabaticity conditions $ |\alpha/(2\pi) | \tau_0 > \omega_{21}$ and $|\alpha / (2 \pi)| < \Omega_R^2$. The physical values applicable to $^{85}$Rb are: the peak Rabi frequency  $\Omega_R = \omega_{21}$ (3.035 GHz), the chirp rate $\alpha/2\pi$=-0.3[$\omega_{21}^2$] (-3 GHz/ns) or faster and the pulse duration $\tau_0 \geq 5.5 \omega_{21}^{-1}$, ($\geq$ 1.8 ns). The one-photon detuning $\Delta$ on the order of the peak Rabi frequency does not diminish adiabatic passage and still leads to the inversion between hyperfine states. The 4-level system may be approximated by an effective three-level $\Lambda$ system giving a qualitatively similar quantum yield when the pulse duration and the chirp rate satisfy $ |\alpha/(2\pi) | \tau_0 \gg \omega_{43}$ and  $\Omega_R \gg \omega_{43}$. If these conditions are not met, the validity of the three-level system breaks down.

This research was supported in part by the National Science Foundation under Grants No. NSF PHY11-25915 and No. NSF PHY12-05454. The work at the University of Connecticut was supported by the Chemical Sciences, Geosciences and Biosciences Division, Office of Basic Energy Sciences, Office of Science, U.S. Department of Energy.


\begin{thebibliography}{8}

\bibitem{To09} D. Tong, S.M. Farooqi, E.G.M. van Kempen, Z. Pavlovic, J. Stanojevic, R. Cote, E.E. Eyler, and P.L. Gould, Phys. Rev. A {\bf 79}, 052509 (2009);  N. Thomas-Peter,  B.J. Smith, A. Datta,  L.  Zhang, U. Dorner,  I.A. Walmsley, Phys. Rev. Lett.  {\bf 107,} 113603 (2011).

\bibitem{Me99} H.J. Metcalf and P. van der Straten, {\em Laser cooling and trapping,}  Springer - Verlag New York, Inc., 1999.

\bibitem{Ni00} M.A. Nielsen and I.L. Chuang, {\em Quantum computation and quantum information,}  Cambridge University Press, 2000.

\bibitem{Ye12} Special issue: {\em Ultracold molecules,} D. Jin and J. Ye, Eds., Chem. Rev. {\bf 112}(9), 2012.

\bibitem{Pa10} V. Patel, V. Malinovsky, S. Malinovskaya, Phys. Rev. A {\bf 81,} 063404 (2010);  G.P. Djotyan, J.S.  Bakos,  G. Demeter, P.N. Ignacz,
 M.A. Kedves, Zs. Soerlei, J. Szigeti, Z.L. Toth, Phys. Rev. A {\bf 68,}
053409 (2003);  E. Heesel, B.M. Garraway, J.P. Marangos, J. Chem.
Phys. {\bf 124,} 024320 (2006).

\bibitem{Ra87}  E.L. Raab, M. Prentiss, A. Cable, S. Chu, and D.E. Pritchard, Phys. Rev. Lett. 59, 2631 (1987).

\bibitem{Ma07}  A. Marian, M. C. Stowe, J. R. Lawall,  D. Felinto,
 J. Ye, {\em Science} \textbf{306,} 2063 (2004); V. Gerginov, C. E. Tanner, S. A. Diddams,  A. Bartels, L. Hollberg,
Opt. Lett. \textbf{30,} 1734 (2005).

\bibitem{Ye08}  M. C. Stowe, A. Pe'er, J. Ye, Phys. Rev. Lett.
\textbf{100,} 203001 (2008).

\bibitem{Dj06}  G.P. Djotyan,  J.S. Bakos, G. Demeter,  P.N. Ignacz,
 M.A. Kedves,  Zs. Soerlei, J. Szigeti, Z.L. Toth, Phys. Rev. A \textbf{68},
053409 (2003).

\bibitem{Co-1-12}  T. A. Collins and S. A. Malinovskaya. Opt. Lett.  \textbf{37}, 2298 (2012); T. A. Collins and S. A. Malinvoskaya. J. Mod. Opt. {\bf 60}, 28 (2013).

\bibitem{key-1}  D. A. Steck, Rubidium 85 D Line Data. http://steck.us/alkalidata
(revision 2.1.4, 2010);  D. A. Steck, Rubidium 87 D Line Data. http://steck.us/alkalidata
(revision 2.0.1, 2008).

\bibitem{Go07}  C.E. Rogers III,  M.J. Wright,  J.L. Carini,  J.A. Pechkis,  P.L. Gould, J. Opt. Soc. Am. B {\bf 24,} 1249 (2007).

\end{thebibliography}
\end{document}